\documentclass[aps,prc,reprint,superscriptaddress,floatfix,twocolumn,nofootinbib]{revtex4-2}
\usepackage{epsfig,rotating,graphics}
\usepackage{amsmath}
\usepackage{amsfonts}
\usepackage{amssymb}
\usepackage{amsthm}
\usepackage{caption}
\usepackage{fontenc}
\usepackage{graphicx}
\usepackage{graphics}
\usepackage{blindtext}
\usepackage{comment}
\usepackage{tikz}
\usepackage{soul}
\usepackage[normalem]{ulem}
\def\be{\begin{eqnarray} &&} 
\def\nonu{\nonumber \\ &&} 
\def\ee{\end{eqnarray}} 

\def\sumint{\int \! \!\ \! \! \! \! \!\ \! \! \!\! \!\sum}

\newcommand{\bfm}[1] {\mbox{\boldmath{$#1$}}}
\newcommand{\bfi}{\begin{figure}}
\newcommand{\efi}{\end{figure}}
\newcommand{\delete}[1]{{\blue{\sout{#1}}}}
\newcommand{\replace}[2]{\red{\sout{#1}}\blue{#2}}

\newcommand{\red}[1]{\textcolor[rgb]{1,0,0}{#1}}

\newcommand{\green}[1]{{\textcolor[rgb]{0,0.7,0.}{#1}}}

\newcommand{\blue}[1]{{\textcolor[rgb]{0,0,1}{#1}}}
\newcommand{\pink}[1]{{\textcolor[rgb]{1,0,0.5}{#1}}}

\newcommand{\gianni}[1]{{ \green{{\bf Gianni}:\ {#1}}}}

\newcommand{\matteo}[1]{{ \pink{{\bf Matteo}:\ {#1}}}}

\begin{document}

\title{$^3$He spin-dependent structure functions within the relativistic Light-Front Hamiltonian dynamics}

\author{Eleonora Proietti}
\affiliation{Dipartimento di Fisica e Geologia. Universit\`a  degli studi di Perugia, INFN Sezione di Perugia.}
\affiliation{Dipartimento di Fisica E. Fermi. Universit\`a di Pisa and INFN Sezione di Pisa, Italy}

\author{Filippo Fornetti}
\affiliation{Dipartimento di Fisica e Geologia. Universit\`a  degli studi di Perugia and INFN Sezione di Perugia, Italy}

\author{Emanuele Pace}
\affiliation{Universit\`a di Roma Tor Vergata, 
Via della Ricerca Scientifica 1, 00133 Rome, Italy}

\author{Matteo Rinaldi}
\email[Corresponding Author: ]{matteo.rinaldi@pg.infn.it}

\affiliation{Istituto Nazionale di Fisica Nucleare
Sezione di Perugia, Via A. Pascoli, Perugia, Italy}

\author{Giovanni Salm\`e}

\affiliation{Istituto  Nazionale di Fisica Nucleare, Sezione di Roma, Piazzale A. Moro 2,
00185 Rome, Italy}

\author{Sergio Scopetta}
\email[S. Scopetta is presently Science Counsellor at the Italian Embassy in Spain, calle Lagasca 98, 28006 Madrid, Spain.]{  sergio.scopetta@esteri.it}
\affiliation{Dipartimento di Fisica e Geologia. Universit\`a  degli studi di Perugia and INFN Sezione di  Perugia, Italy}

%\submitto{\PS}
%\maketitle

\date{\today}
\begin{abstract}
 $^3$He spin-dependent structure functions, $g^3_1(x)$ and $g^3_2(x)$, which parametrize
 the hadronic tensor in polarized deep-inelastic scattering, were evaluated within the Poincar\'e covariant light-front framework. {The Bakamjian-Thomas construction of the Poincar\'e generators allows us to make use of a realistic
{$^3$He} wave function, obtained from {refined} nuclear phenomenological potentials. The same approach was already {successfully applied to the $^3$He and $^4$He unpolarized deep-inelastic scattering.}  To investigate the neutron polarized structure functions, $g^n_1$ and $g^n_2$,}
a {readily implementable} procedure, aimed at extracting the neutron spin structure functions from those of $^3$He,  is shown to hold.  Moreover, the first moment of  $g^3_1(x)$  was evaluated, aiming at  providing a valuable check of the Bjorken sum rule. The present analysis is relevant for experiments involving polarized beams planned at the future facilities, like the Electron Ion Colliders.

\end{abstract}

\maketitle

 %Uncomment for keywords
%\noindent{\it Keywords}: Polarised Deep inelastic lepton scattering, Poincar\'e covariance, light-front Hamiltonian dynamics
%, light-nuclei

\section{Introduction}
\label{Sect_intro}
Deep-inelastic scattering (DIS) involving polarized leptons off polarized nuclear targets are the key to accessing unique information on the  spin structure inside nucleons  and  nuclei (see,  e.g., Ref. \cite{Lampe:1998eu} for an introduction). One of the main goals of the polarized DIS  is the flavor decomposition of the nucleon spin-dependent structure functions (SSFs), $g^N_1$ and $g^N_2$, {by making available both the proton and the neutron ones}. Recalling that in nature a free-neutron target is lacking, one has to resort to a polarized nuclear target, e.g. deuteron or  $^3$He targets.

The use of a nuclear target for addressing the underlying nucleonic degrees of freedom (dofs) requires a reliable  description of the nuclear effects, in order to minimize the model dependence in the extraction of the neutron  structure functions from the experimental data.  Due to its spin structure, the polarized $^3$He can be considered as an effective polarized neutron target (see, e.g.,  Refs. \cite{Friar:1990vx,CiofidegliAtti:1993zs,Schulze:1992mb,DelDotto:2017jub,Rinaldi:2012ft,Pace:2001cm}, 
for the proposal of extraction of  neutron  structure functions and parton distributions from those of $^3$He). Therefore,  a  realistic description of the polarized three-nucleon bound system that takes into account as many general principles as possible has to be implemented, as described,  { for the unpolarized case}, in Refs. \cite{Pace:2022qoj,Pace:2020ned} and references quoted therein.   

In general,  the DIS asymmetries of polarized electrons from polarized nuclear targets (mainly ${\rm J}=1/2$ and ${\rm J}=1$ nuclei) contain the  nuclear spin-dependent structure functions, $g_1^A$ and $g_2^A$, with $A$ identifying the target nucleus  (see Refs. \cite{Friar:1990vx,Woloshyn:1988nd,Khan:1991qk,CiofidegliAtti:1993zs, Kaptari:1994di,Kulagin:1994cj,Piller:1995mf,Melnitchouk:1994tx,Schulze:1997rz,Bissey:2000ed,Bissey:2001cw,Sargsian:2001gu,Afnan:2003vh,Khanpour:2017fey} for a theoretical overview). These scalar functions allow one  to parametrize the antisymmetric part of the hadronic tensor, after imposing the symmetry requirements. Given the relevant information they convey, a huge effort aimed at measuring these nuclear SSFs for a polarized $^3$He target was made by different experimental collaborations, e.g. at SLAC, {DESY} and JLAB \cite{E142:1993hql,E142:1996thl,E154:1997xfa,HERMES:1997hjr,JeffersonLabHallA:2003joy,JeffersonLabE94-010:2003dvv,JeffersonLabHallA:2004tea,JeffersonLabHallA:2014mam,JeffersonLabHallA:2016neg} (see, e.g., Refs. \cite{E143:1998hbs,E155:1999pwm,HERMES:2006jyl}  for a polarized deuteron target).

Already many years ago, it was {recognized} within a  framework {based on a non Poincar\'e-covariant $^3$He spectral function} \cite{CiofidegliAtti:1993zs} that the neutron polarized structure function 
%\gianni
{in the Bjorken limit}, $g_1^n(x)$, can be obtained from the $^3$He spin structure function, $g_1^3(x)$, {by using}
\be
\label{eq:gn1}
    \bar g_1^n(x)= \dfrac{1}{p_1^n} \big[g_1^3(x)- 2 p_1^p g_1^p(x)  \big]~,
\ee
where $p_1^p$ and $p_1^n$ are the effective proton and neutron polarizations \cite{CiofidegliAtti:1993zs}. This procedure was actually largely used by the experimental collaborations (see, e.g., Ref. \cite{E142:1996thl}). 
However, energy and  momenta involved in these processes are large and a non Poincar\'e-covariant framework could not be the proper one.

The aim of this letter is to study whether Eq. \eqref{eq:gn1} still holds in the Poincar\'e-covariant formalism developed in Ref. \cite{Pace:2020ned} within the light-front Hamiltonian dynamics (LFHD) \cite{Dirac:1949cp,KP}. It is worth noticing that LFHD  is ideal for  describing  
{electromagnetic (em) processes at high energies, as well as}  
%\replace{strong interacting}
{bound} systems with a finite number of dofs.
%involved 
%\delete{in  electromagnetic (em) processes, at high energies}. 
This formalism  coupled to  the Bakamjian-Thomas (BT) construction of the Poincaré generators \cite{Bakamjian:1953kh} allows one to easily introduce the standard nuclear interactions, which fulfill  the proper commutation rules. %\gianni
Hence,  
one can  use the  sophisticated phenomenology of nuclear physics within a Poincaré-covariant approach.

To evaluate the $^3$He SSFs, we adopt the realistic wave function of Ref. \cite{Kievsky:1992um}, obtained from the  phenomenological {charge-dependent} potential Argonne v18  (Av18) \cite{Wiringa:1994wb}, {widely used to achieve a very  accurate description of  nuclear systems} 
%\gianni
{(as shown in Ref. \cite{Pace:2022qoj}, the effects of the three-nucleon forces {are marginal}).}
Such an approach was already applied  i) to $^3$He for evaluating both  spectral function \cite{DelDotto:2016vkh,Pace:2020ned} and  {nuclear} transverse momentum-dependent parton distribution functions (TMDs) \cite{Alessandro:2021cbg} and, recently, ii)  to $^3$He, $^3$H and $^4$He nuclei for  estimating the EMC effect \cite{Pace:2022qoj,Fornetti:2023gvf}. 
Let us also remind that {our investigation, based on needed general symmetries and refined nuclear description,} will be 
%\replace{extremely relevant for the program of}
{a well-grounded theoretical framework for the analysis of data gathered at}   the future Electron Ion Colliders \cite{AbdulKhalek:2021gbh,Anderle:2021wcy}, where polarized $^3$He nuclei (and deuterons) will be used. 

{\it Polarized DIS off nuclei. }
Since the goal of the present analysis is to extend the light-front (LF) approach
%, already applied  to describe the $^3$He spectral function \cite{Pace:2020ned} and to evaluate the EMC effect for $A=3,4$ nuclei \cite{Fornetti:2023gvf,Pace:2022qoj}, 
to the evaluation of the $^3$He SSFs, we {shortly} recall the  formalism describing the  polarized {inclusive cross-section in  DIS processes.}
% \gianni{In the laboratory frame,} 
 The inclusive inelastic scattering of polarized leptons by a polarized { spin 1/2} target $A$, i.e. $\vec{\ell}+\vec A=\ell' +X$ %\gianni{i leptoni sono polarizzati } 
 reads  in one-photon exchange
approximation { in the laboratory system} as follows {(see, e.g., Ref. 
%\gianni
 {\cite{Blankleider:1983kb,CiofidegliAtti:1994cm} } 
%{for the analogous quasi-elastic case} 
\be
\frac{d\sigma( +S)}{d\Omega_2 d\nu }-\frac{d\sigma( -S)}{d\Omega_2 d\nu } =
4\frac{\alpha_{em}^2}{ Q^4 }~m^2_e
\frac{{\cal E}'}{{ \cal E}}
\nonu
\times
 L^{a,\mu\nu}(h_\ell)  W_{a,\mu \nu}^{A}(P_A,{\bf S},{\cal M},T_{Az}, q)~,
 \label{eq:crosa-1}
   \ee
%\emanuele{ (see, e.g., \cite{Barone:2001sp}) 
%\be
%\frac{d\sigma( +S)}{d\Omega_2 d\nu }-\frac{d\sigma( -S)}{d\Omega_2 d\nu } = \frac{\alpha_{em}^2}{2 Q^4 M_A}~
%\frac{{\cal E}'}{{ \cal E}}
%\nonu \times~L^{a,\mu\nu}
%{(h_\ell) } W_{a,\mu \nu}^{A}(P_A,{\bf S},{\cal M},T_{Az}, q)~,
% \label{eq:crosa-1e}
%  \ee}
where
$k^\mu=({\cal E}, {\bf k})$ and
$k'^{\mu}=({\cal E'}, {\bf  k}')$  are incoming and outgoing charged leptons 4-momenta, respectively, the square 4-momentum transfer in
 ultrarelativistic approximation is 
$Q^2 =-q^2= -(k-k')^2 =| {\bf q}|^2 - \nu^2=4 {\cal E}
{\cal E}' \sin^2(\theta_{\ell}/ 2)$ (with
${\bf q }= {\bf k} - {\bf k}'$, $\nu= {\cal E} - {\cal E}' $ and
$\theta_{\ell}\equiv \theta_{\widehat{{\bf k}  {\bf  k}'}}$). 

The antisymmetric leptonic tensor $L^a_{\mu\nu}$
%,  calculable in QED, 
has the form
\be
L^a_{\mu\nu}(h_\ell)=
 ih_\ell \varepsilon_{\mu\nu\alpha\beta}~ \frac{k^\alpha~q^\beta}
 %\over 
 {2 m^2_e}~,
\label{lmunua}
\ee
%\emanuele{\be
%L^a_{\mu\nu}(h_\ell)=2 ih_\ell \varepsilon_{\mu\nu\alpha\beta}~ k^\alpha~q^\beta~,
%\label{lmunua-e}
%\ee}
where $ h_\ell $ is the helicity of the incoming electron 
%\gianni
{(in the ultrarelavistic limit $h_\ell=\pm 1$)} and 
$\varepsilon_{\mu\nu\alpha\beta}$
the fully antisymmetric  Levi-Civita tensor ($\varepsilon_{0123}=-1$). %\gianni{la convenzione di solito \`e opposta.....}).
In Eq. \eqref{eq:crosa-1},  $W^{A}_{a,\mu \nu}(P_A,{\bf S},{\cal M},T_{Az}, q)$ is the antisymmetric part of the hadronic tensor for inclusive {lepton DIS } off an $A$-nucleon nucleus with nuclear LF ground state 
 $ |\Psi_{0}; {\bf S}, {\cal M},T_{Az};P_A\rangle_{LF}$. Let us assume that the nucleus: $i)$ is polarized along ${\bf S}$ 
 %\red
 {(belonging to the scattering plane)}, with spin projection $\cal M$, $ii)$ has an isospin third-component $T_{Az}$, and $iii)$ total 4-momentum $P_A \equiv [P_A^-, P_A^+,{\bf P}_{A\perp}]$, 
 %\gianni{notice that in a collider the lab frame does not coincide with the rest frame, we have to choose the formalism with this in mind, below the rest frame has been introduced.}, 
 with $ P_A^2 = M_A^2$. 

  In the polarized case the antisymmetric part 
%\delete{of this tensor} 
can be described {at {finite} $Q^2$} in terms of %\replace{the}
{two} {\em nuclear} SSFs,
{$G^A_1(x;Q^2)$ and  $G^A_2(x;Q^2)$}, namely
 %, respectively
 {\cite{CiofidegliAtti:1994cm,Ellis:1973kp}
 \be
 W^{A}_{a,\mu \nu}  =  i ~ %\frac
 {\epsilon_{\mu \nu \rho \sigma} q^\rho}
% \over 
% {P_A\cdot q}
 ~\Biggl \{  S^\sigma %P_A\cdot q~
 ~\frac{G^A_1 (x,Q^2)}
% \over 
 {M_A} \nonu
+ \left [ S^\sigma -
\frac{S \cdot q }  {P_A \cdot q} P_A^\sigma \right ] ~[{P_A \cdot q}]~\frac{G^A_2 (x,Q^2)}
%\over
{M^3_A}   \Biggr \} ~ ,
\label{hta}
\ee}
%\emanuele{ \be W^{A}_{a,\mu \nu}  = \frac{2 i ~ M_A \epsilon_{\mu \nu \rho \sigma} q^\rho} { P_A \cdot q }
% \nonu \times\left \{  S^\sigma  g^A_1 (x,Q^2) + \left [ S^\sigma - \frac{S \cdot q }  {P_A \cdot q} P_A^\sigma \right ] g^A_2 (x,Q^2)   \right \} \ ,
%\label{hta-e}
%\ee}
where in the rest frame %\gianni{prima si parla di Lab frame!!!}:
$S = (0,{\bf S})$ with $|{\bf S}|=1$, $P_A^+ = P_A^- = M_A, ~ {\bf P}_{A\perp}=0$  and ${\bf q} \equiv (0,0, q_z=-|{\bf q}|)$.
Furthermore, in this reference frame {one has ${\bf k}_\perp={\bf k}^\prime_\perp$ and let us assume   $k_y = k'_y =0$.}   %\gianni{non capisco assume, Forse Since $\hat{\bf q}\equiv \hat{\bf e}_z$ then ${\bf k}_\perp={\bf k}^\prime_\perp$}.

{In terms of $G^A_1 (x,Q^2)$ and $G^A_2 (x,Q^2)$, the spin  structure} {functions $g_1^A(x,Q^2)$ and $g_2^A(x,Q^2)$, that we are going to evaluate and that are expected to scale in the Bjorken limit, are \cite{Ellis:1973kp}}
\be
\hspace{-.5cm} g_1^A(x,Q^2) = P_A\cdot q~ \frac{G^A_1 (x,Q^2)}{M^2_A}
\nonu
\hspace{-.5cm} g_2^A(x,Q^2) = [P_A\cdot q]^2  ~ \frac{G^A_2 (x,Q^2)}{M^4_A} 
\label{eq:}
\ee

{\it Polarized LF spectral function.} To evaluate the nuclear SSFs,
as already discussed in Refs. \cite{Pace:2020ned,Pace:2022qoj,Fornetti:2023gvf}, one  assumes  the impulse approximation (IA) framework, where a photon, with high virtuality,
{is absorbed by} a nucleon in the nucleus leaving a fully interacting $(A-1)$-nucleon system as a spectator. 
We remind that, in the chosen frame, the IA
%\gianni
{is based  on  free em current operators and}  fulfills the Poincaré covariance \cite{Lev:1998qz}.
In this 
{framework}, the main nuclear ingredient affecting  the nuclear structure functions {(both unpolarized and polarized),} is the spin-dependent LF spectral function,  
{given by a $2\times 2$ matrix} \cite{Pace:2020ned,DelDotto:2017jub,Alessandro:2021cbg}.
%\replace{We recall that this quantity }
{Its trace} is the {probability distribution to find inside a bound system
 a particle with a given
3-momentum when 
%\replace{the rest of the}
{the spectator} system has a given energy $\epsilon$.} 
  For a nucleus, 
  % \replace{polarized  along the }
   {with} polarization vector ${\bf S}$, the corresponding LF spectral functions reads \cite{Pace:2020ned,DelDotto:2017jub,Alessandro:2021cbg}
\be
 {\cal P}^{\tau}_{\sigma'\sigma}(\tilde{\bfm \kappa},\epsilon,{\bf S},{\cal M}) = \rho(\epsilon) 
\nonu 
\times
 \sum_{J J_{z}\alpha}\sum_{T t } ~
_{LF}\langle  t,T ; 
\alpha,\epsilon ;J J_{z}; \tau\sigma',\tilde{\bfm \kappa}|{\psi}_{\cal{J}\cal{M}};{\bf S}, T_A T_{Az}\rangle
\nonu
\times ~ \langle {\bf S}, T_A T_{Az};
{\psi}_{\cal{J}\cal{M}}|\tilde{\bfm \kappa},\sigma\tau; J J_{z}; 
\epsilon, \alpha; T, t\rangle_{LF} ~ , 
\label{LFspf}
\ee
where  $\rho (\epsilon)$ is the energy density of the $(A-1)$-nucleon state and $|{\psi}_{\cal{J}\cal{M}};{\bf S}, T_A T_{Az} \rangle$ is the $A$-nucleon intrinsic   
  ground state, with total angular momentum ${\cal J}$ (third component ${\cal M}$), isospin $T_A$ (third component $T_{Az}$) and rest-frame polarization 
  $S\equiv\{0, {\bf S}\}$.
  %\begin{align}
%&  {\cal P}^{\tau, m',m}_{\sigma'\sigma}(\tilde{\bfm \kappa},\epsilon,\hat z) =
%\rho(\epsilon) ~\sum_{J J_{z}\alpha}\sum_{T t } ~
%\\
%\nonumber 
%& \times ~
%_{LF}\langle  t T ; 
%\alpha,\epsilon ;J J_{z}; \tau\sigma',\tilde{\bfm %\kappa}|
%j,j_z=m';
%\epsilon^A,\Pi; T_A T_{Az}\rangle 
%\\
%\nonumber
%& \times ~
%\langle T_A T_{Az}; \Pi,\epsilon^A; j,j_z=m; 
%|\tilde{\bfm \kappa},\sigma\tau; J J_{z}; 
%\epsilon, \alpha; T t\rangle_{LF} 
%\label{SFex}
%\end{align}
%with  $\rho (\epsilon)$ the energy density of the $(A-1)$-nucleon state.
%\delete{Details on the definition and properties of the above quantity can be found in {Pace:2020ned,DelDotto:2016vkh,Alessandro:2021cbg}.}
%This expression is valid for a nucleon in a nucleus 
% of total 4-momentum $P_A$ in the laboratory frame, with LF momentum components 
%  $[P_A^-, {\blf P}_A] \equiv [P_A^-, P_A^+,{\bf P}_{A\perp}]$.
%The state
%$|j,j_z=m; \epsilon^A, \Pi; T_A T_{Az}\rangle$ is the intrinsic ground state of the $A$-nucleon system, with angular momentum $(j,j_z)$, energy $\epsilon^A$, parity $\Pi$, isospin  $(T_A T_{Az})$, and polarization along $\hat z$  with LF total energy $P_A^-$. 
The quantum numbers ($J, J_{z}; 
\epsilon;  T, t$) describe the angular momentum, intrinsic energy and isospin of the fully-interacting $(A-1)$-nucleon system, respectively; while $\alpha$ indicates the other quantum numbers needed to completely identify this system.
 The 
 %\delete{intrinsic} 
 state $\left |\tilde{\bfm \kappa}, \sigma \tau; J J_{z}; 
\epsilon, \alpha, T, t\right\rangle_{LF}$ %\replace{describes the cluster, composed by}
{is the Cartesian product of} the
fully-interacting intrinsic state of the 
 $(A-1)$ spectator system times
 a free-nucleon state with  
spin $\sigma$,  isospin $\tau$ and moving
 in the intrinsic reference frame of the whole cluster $[1, (A-1)]$ 
with  intrinsic momentum $\tilde{\bfm \kappa}\equiv (\kappa^+,{\bfm \kappa}_{\perp})$ (cfr. Refs. \cite{Pace:2020ned,Alessandro:2021cbg}) 
%\delete{of a nucleon in the reference frame of the  cluster $[1, (A-1)]$ is different from the momentum $\tilde{\bfm p} \equiv ( p^+,{\bf p}_{\perp})$ in the laboratory frame}.
In {particular}, $\kappa^+ = \xi \mathcal{M}_0(1,A-1) $ {and $\kappa_\perp = p_\perp - \xi P_{A\perp}$}, where $\xi= p^+/P^+_A$, with $(p^+,{\bf p}_\perp )$ the nucleon momentum in the {rest} frame, %\gianni{dobbiamo ricordare che abbiamo usato anche rest frame...} 
and $\mathcal{M}_0(1,A-1)$ the free mass of the $[1,A-1]$ cluster. 
The latter depends on the total energy of the spectator system: 
$E_S=\sqrt{M_S^2+ |{\bf \kappa}_\perp|^2+\kappa_z^2}$ (see Ref. \cite{Alessandro:2021cbg} for details) {with} $M_S = (A-1)m+\epsilon$.
As noticed in \cite{DelDotto:2016vkh}, the description of the cluster $[1, A-1]$ through the {state}
 $\left |\tilde{\bfm \kappa}, \sigma \tau; J J_{z}; 
\epsilon, \alpha; T, t\right\rangle_{LF}$ fulfills the macrocausality \cite{KP}. %\replace{}{ MA SERVE ??? %Finally,  the unpolarized  spectral function is given by the following trace \delete{the spin-dependent one }\cite{DelDotto:2016vkh}:
%\be
%  {\cal P}^{\tau}(\tilde{\bfm \kappa},\epsilon)  = ~ %\frac{1}  {2~ {\cal{J}} + 1} ~ {1 \over 2}~ \sum_{\cal{M}}  \sum_\sigma ~ {\cal P}^{\tau}_{\sigma\sigma}(\tilde{\bfm \kappa},\epsilon,{\bf S},{\cal M})
%{\cal P}^{\tau,m,m}_{\sigma\sigma} (\tilde{\bfm \kappa},\epsilon,\hat z).
%\label{unpol}
%\ee
%\gianni{Notice that the average on the nucleus polarization is dummy.}}

{\it Spin-dependent nuclear SF in IA.} In analogy with the strategy developed for the unpolarized case, one can show that the IA description of the reaction mechanism for polarized DIS leads to a convolution formula between the nuclear LF spectral function and  the nucleon hadron tensor, without 
any { dynamical} modification of the nucleon internal structure \cite{Pace:2020ned,Pace:2022qoj,DelDotto:2016vkh}. 
%\gianni
{Hence, the antisymmetric hadronic tensor is given by} 
\be
 W^{a,\mu \nu}_A = 
 \sum_N \sum_{\sigma \sigma'}\sumint {d\epsilon 
}
  \int \frac{d{\bfm \kappa}_{\perp} ~d\xi }{2~(2 \pi)^3 ~  \kappa ^+}~ \frac{1} {\xi} \dfrac{E_S}{(1-\xi)}
  \nonu
  \times
 {\cal P}^{N}_{\sigma\sigma'}(\tilde{\bfm \kappa},\epsilon,{\bf S},{\cal M})  
 ~ w^{a,\mu \nu}_{N,\sigma',\sigma}({p},{q})~,  
\label{HTTT}
\ee
where $w^{a,\mu \nu}_{N,\sigma',\sigma}({p},{q})$ is the antisymmetric part of single-nucleon  hadronic tensor (see  Sect. S1  in  supplemental material for details).
The polarised deep-inelastic nuclear SSFs $g_1^A$ and $g_2^A$ can be obtained from specific combination of the hadronic tensor components, namely
{ \be
g_j^A(x, Q^2) = (-1)^j ~ \frac{i ~  \nu}  {2 |{\bf q}|} \left [ \dfrac{Q^2}{|{\bf q}|^2}\dfrac{W_{A,02}^a}{S_x}  -
\dfrac{\nu}{|{\bf q}|} \dfrac{W^a_{A,12}}{S_z}
\right.
\nonu
-\left. \dfrac{W_{A,02}^a}{S_x} \dfrac{1+(-1)^j  }{2}
 \right ]~,
 \label{HT55}
 \ee
 with $j=1,2$. }

%\gianni
{Substituting in Eq. \eqref{HT55} the  expression of $W^{a,\mu\nu}_A$ given in Eq. \eqref{HTTT}, one gets  the IA results for the nuclear SSFs (see supplemental material for details).}
%\delete{Consequently to Eq. (\ref{HT55}), and analogously to the unpolarized case {Pace:2022qoj}, the nuclear SSFs are given by a convolution between a nuclear {distribution} and the corresponding  free nucleon SSFs. As already discussed for the $F_1^A$ and $F_2^A$ structure functions \cite{Pace:2022qoj}, 
  In our Poincar\'e covariant framework the spectral function of the nucleus has  non trivial dependences on the removal energy of the spectator system and the free-nucleon momentum.
Therefore its calculation is numerically challenging. 
However, {as discussed in \cite{Pace:2022qoj} a very effective approximation is to perform our evaluations in the Bjorken limit,} given  the large values of  momentum transfer involved in the experiments. 
Within this approximation, the integrals on $\xi$  and ${\bf \kappa}_\perp$ in Eq. (\ref{HTTT}) commute, and, in turn, both of them commute with the one on the removal energy $\epsilon$.
This simplification allows one to  
 evaluate the nuclear SSFs from the {transverse momentum dependent distributions (TMDs)} (see Ref. \cite{Pace:2022qoj} and supplemental material):
%\gianni{Si \`e nel limite di BJ 
\be
\label{eq:gj}
 g_j^A(x)= g_j^{A,n}(x)+g_j^{A,p}(x)= \sum_{N=n,p} \int_{\xi_{min}}^1 \hspace{-0.2cm}d\xi 
\nonu 
\times~\Big\lbrace g_1^N(z) l^N_j(\xi)+g_2^N(z) h^N_j(\xi)   \Big\rbrace~, 
%j=1,2
\ee
%\blue{ Da cancellare\be
%\label{eq:gjb}
 %g_j^A(x;Q^2)= g_j^{A,n}(x;Q^2)+g_j^{A,p}(x;Q^2)= \sum_{N=n,p} %\int_{\xi_{min}}^1 \hspace{-0.2cm}d\xi 
%\nonu 
%\times~\Big\lbrace g_1^N(z;Q^2) l^N_j(\xi)+g_2^N(z;Q^2) h^N_j(\xi)   %\Big\rbrace~, 
%j=1,2
%\ee}
where
 %\matteo
 {$\xi_{min}=x m/M_A$, $z=mx/\xi M_A$ and} $h^N_1(\xi) =0$.
 The quantities $l_j^N(\xi)$ and $h^N_j(\xi)$ can be related to the  TMDs
 {of the nucleons in $^3$He}, {defined in  Ref. \cite{Alessandro:2021cbg}.}
{Explicit expressions of the nuclear SSFs in terms 
%of the  
of 
the free nucleons ones and the nuclear TMDs can be found in supplemental-material Sect. S2, as well as the complete definitions of the light-cone momentum distributions 
$l_j^N(\xi)$ and $h^N_j(\xi)$.}
%introduced above
 %\gianni{gi\`a detto :Details on the explicit relations between these quantities and the TMDs, can be found in the SM}. 
 %\delete{From now on we omit the implicit $Q^2$ dependence of the SSFs, being applied the Bjorken limit.}\gianni{ si \`e dichiarato che si lavora nel limite di BJ}

{\it Results.} 
In order to evaluate Eq. (\ref{eq:gj}), 
one needs to properly choose a parametrization for the free-nucleon SSFs. In the present investigation we 
use the {standard-scenario NLO parametrization}  of  {Tab. I in} Ref. \cite{Gluck:2000dy}   for $g^N_1(x)$ {(GRSV parametrization)}.
The SSF  
$g^N_2(x)$ {can be} given in terms of $g_1^N$  following the well-known Wandzura-Wilczek {(WW)}  { approximation}  \cite{Wandzura:1977qf}, i.e.,
 \be
 \label{WW}
 g_2^N(x)=- g_1^N(x)+\int_x^1 dy~\frac{g_1^N(y)} 
 {y}~.
 \ee
 We point out that the evaluation of $g_1^3(x)$ 
 { in the Bjorken limit}
 is not affected by the choice of $g_2^N$
since $h_1^N(\xi)=0$ in Eq. (\ref{eq:gj}).

%In order to obtain  the numerical evaluation of the $^3$He SSFs, we  adopted the free-nucleon SSFs of Ref. \cite{Gluck:1991ng} 
$^3$He TMDs
 %,  presented in Ref. \cite{Pace:2022qoj}.
 were  obtained by using the
{refined} $^3$He wave function of Ref. \cite{Kievsky:1992um}, which corresponds to the phenomenological Av18 potential \cite{Wiringa:1994wb}.

{For an easy comparison with the neutron SSF, $g^3_1(x)$ and $g^3_2(x)$ are multiplied by the factor $m/M_3$, as we already did in \cite{CiofidegliAtti:1994cm} (see also \cite{Schulze:1992mb}). } In Fig. \ref{fig:g1g2} we compare our predictions for $g_1^3(x)$ (left panel) and $g_2^3(x)$ (right panel) with present data of Refs. \cite{E142:1996thl,JeffersonLabHallA:2003joy,JeffersonLabHallA:2016neg}. As one can see, {without}    free parameters {and exotic effects (e.g. a component with $\Delta$ dofs)}, the realistic calculations  describe well the data.
In this figure the {non Poincar\'e-covariant} calculations, obtained by applying the expressions of Ref. \cite{CiofidegliAtti:1993zs}, but with the Av18 interaction, cannot be distinguished from the Poincar\'e-covariant results { (see  supplemental material for further details)}.  

 \begin{figure*}[t]
%\begin{center}
\includegraphics[width=8.2cm]{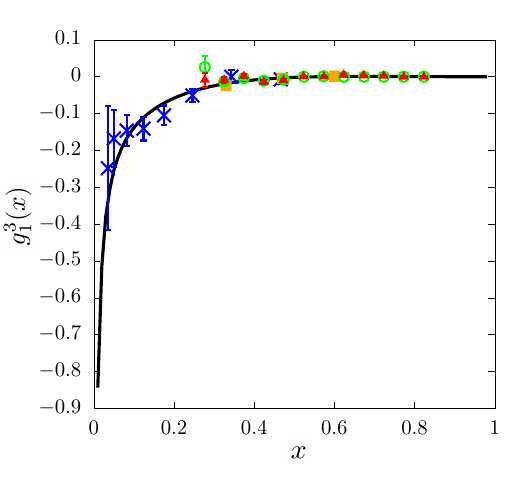}
\includegraphics[width=8.2cm]{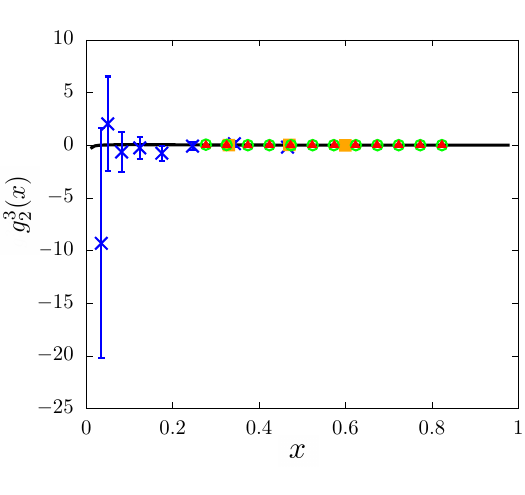}
\caption{ (Color online). The $^3$He SSFs obtained with the Poincaré covariant approach (full line)  vs. the Bjorken variable $x$. The experimental data are taken from 
Ref. \cite{E142:1996thl} (crosses), Ref. \cite{JeffersonLabHallA:2003joy} (squares), and Ref. \cite{JeffersonLabHallA:2016neg}:  {empty circles}   correspond to an incoming-electron energy ${\cal E}=5.89$ GeV and  triangles    to ${\cal E}= 4.74$ GeV). 
Left panel:  $g^3_1(x)$. Right panel:   $g^3_2(x)$.   %\gianni{Io toglierei la curva NR, ma direi nel testo che si ha un overlap} 
}
\label{fig:g1g2}
%\end{center}
\end{figure*}
  The present experimental accuracy can only allow to assess the overall pattern  of both nuclear SSFs, that 
 %however %it 
{appears} to be in nice agreement with 
 the theoretical predictions.

In Fig. \ref{fig:g1g2a}, the full calculation of the $^3$He SSFs are compared 
%\delete{, obtained within the present Poincaré covariant approach, Eq. \eqref{eq:gj}, with the proton and the neutron contributions and}
with the free neutron SSFs {obtained from the GRSV parametrization  \cite{Gluck:2000dy}. 
%\gianni
{For the sake of completeness}, the proton and neutron contributions to $g^3_j(x)$ are also shown. As expected from the $^3$He spin structure, $g_{1(2)}^{3,n}(x)$ is {largely} dominant with respect to $g_{1(2)}^{3,p}(x)$. Moreover, the pattern of $g_{1(2)}^3(x)$ is very similar to that of $g_{1(2)}^n$, and this is consistent  with the idea that $^3$He can be considered as {an effective neutron target}. However, relevant differences can be appreciated for $0.2<x<0.8$ due to the presence of nuclear effects.

{\it Extraction of the neutron SSFs.} {In addition to }the intrinsic relevance of the $^3$He SSFs, particularly for the experimental program of  future EICs, 
{their investigation, both experimental and theoretical, plays a key role for gaining  unique information on the neutron ones}.

{Workable procedures to exploit the $^3$He target for extracting  i)  the  neutron  unpolarized and polarized structure functions \cite{CiofidegliAtti:1993zs,Scopetta:2006ww} and ii) the neutron generalized parton distribution \cite{Rinaldi:2012ft},  were proposed several years ago and extensively used, { in the SSFs case},
for experimental analyses (see e.g. Ref. \cite{E142:1996thl})}.  As already shown in Refs. \cite{Alessandro:2021cbg,Pace:2022qoj}, the $^3$He TMDs are peaked around $\xi \sim 1/3$ and therefore Eq. (\ref{eq:gj}) can be approximated as follows {\em{(see supplemental material for details)}}:
\begin{align}
\label{eq:gapp}
g_j^A(x) \sim \bar g_j^{A}(x) = 2 p^p_j g_j^p(x)+p^n_j g_j^n(x)~, 
\end{align}
where $p^{n(p)}_1$ and  $p^{n(p)}_2$ are neutron (proton) longitudinal and transverse effective polarizations, respectively. They are defined as the  integral over $\xi$ of the {corresponding} light-cone momentum distributions {(see Ref. \cite{Alessandro:2021cbg})}. Numerically, they read {(within parenthesis the  results obtained by using a non relativistic $^3$He spectral function \cite{Alessandro:2021cbg})}
%(for the NR case see \cite{CiofidegliAtti:1993zs})
:
\begin{align}
\label{eq:eff_pol}
    &p_1^n \simeq 0.873 (0.893)~, ~~~p_1^p \simeq -0.0230(-0.0212)
    \\
    \nonumber
         &p_2^n \simeq 0.873 (0.893)~, ~~~p_2^p \simeq -0.0245(-0.0212)~.
\end{align}
%\gianni {5 cifre sono inutili}
We remind that differences between the longitudinal and transverse effective polarizations 
%are measures of relativistic effects 
{are due to relativistic effects}.

{If the approximation of Eq. (\ref{eq:gapp})} works, one can extract the neutron SSFs from those of  $^3$He and  proton, obtaining
\begin{align}
\label{eq:gn}
    \bar g_j^n(x)= \dfrac{1}{p_j^n} \big[g_j^3(x)- 2 p_j^p g_j^p(x)   \big]~~ (j=1,2)~.
\end{align}
\begin{figure*}[h]
\begin{center}
\includegraphics[width=7.5cm]{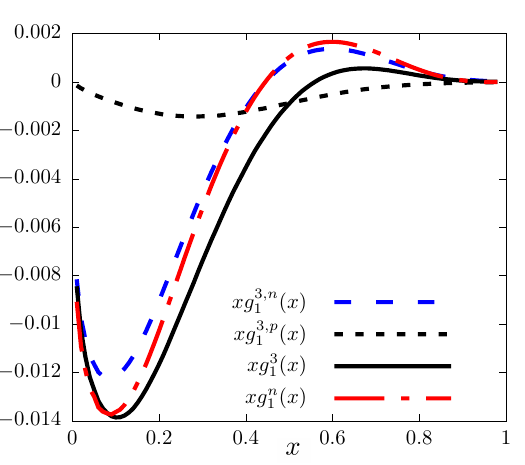}
\includegraphics[width=7.5cm]{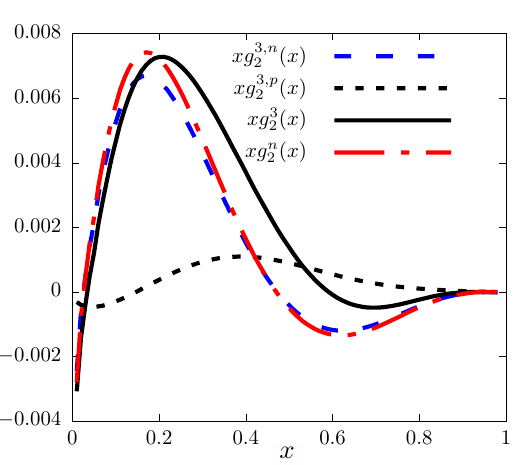}
\caption{ (Color online). Theoretical $^3$He SSFs,  Eq. (\ref{eq:gj}), compared to the corresponding free-neutron ones. Solid lines: full calculations. Dashed lines: neutron contribution to the nuclear SSFs. Dotted lines: proton  contribution to $g^3_j(x)$. Dot-dashed lines: free neutron SSFs,  evaluated by using the GRSV parametrization  \cite{Gluck:2000dy}.  Left panel:  $x g_1$. Right panel:  $x g_2$. }
\label{fig:g1g2a}
\end{center}
\end{figure*}

%\begin{comment}
\begin{figure*}[h]
\begin{center}
\includegraphics[width=7.8cm]{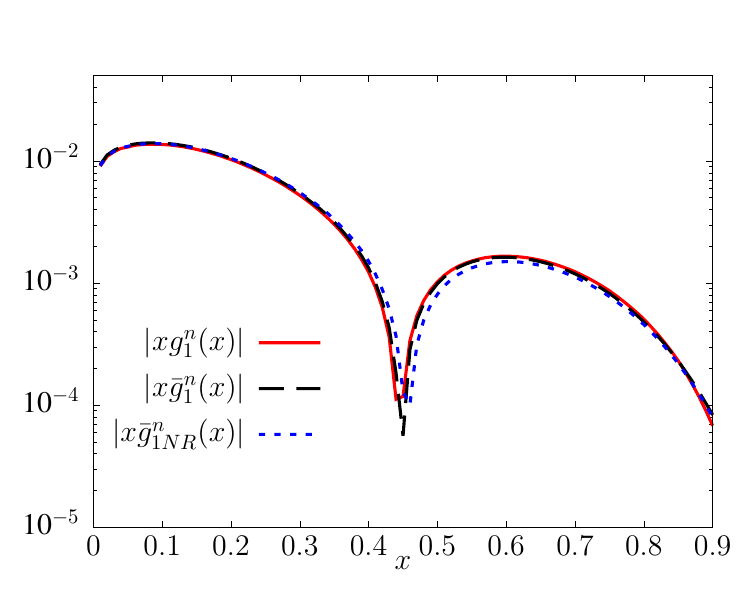} \hskip -0.2cm
\includegraphics[width=7.8cm]{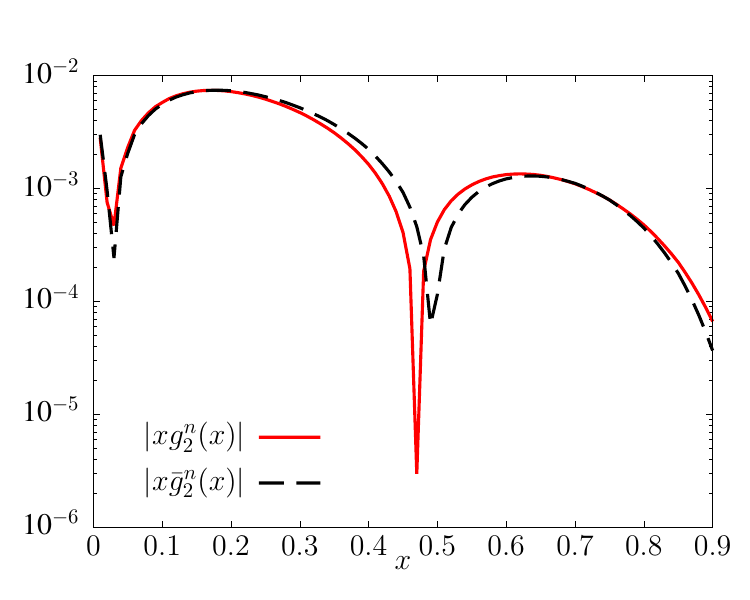}
\caption{ (Color online). Comparison between the {absolute values of the} %full calculations of the $^3$He SSFs Eq. (\ref{eq:gj}) 
 approximation of Eq. (\ref{eq:gn}) for the neutron SSF and the free neutron SSF. Left panel:  $ |x g_1|$. Right panel:   $ |x g_2|$. Solid lines:  neutron SSFs from  the GRSV parametrization  \cite{Gluck:2000dy} Dashed lines:   free neutron SSFs, extracted from Eq. (\ref{eq:gn}). Dotted line:  neutron SSF $\bar g_1^n$ obtained from Eq. (\ref{eq:gn}) using the non-relativistic effective polarizations.} 
\label{fig:3app}
\end{center}
\end{figure*}
%\end{comment}
%%%%%

In Fig. \ref{fig:3app} we numerically test the validity of this approximation by comparing the neutron SSFs obtained from Eq. (\ref{eq:gn}) and the ones  {corresponding to} the (GRSV)} parametrization  \cite{Gluck:2000dy}. Moreover,
 we compare our calculations of $ \bar g_j^n$  with the corresponding ones obtained {by using}  in Eq. \eqref{eq:gn} the non relativistic effective polarizations. 
Due to the large uncertainties in the experimental data for the $^3$He SSFs,  $g_1^3(x)$ in Eq. (\ref{eq:gn}) is simulated
by the full calculation of Eq. (\ref{eq:gj}), 
%\gianni
{which contains the relevant nuclear effects.} {Up to $x = 0.9$ the proposed procedure works very well and somewhat better than the {extraction based on non relativistic polarizations}. Actually the difference between the model used for $g_j^n(x)$ and $\bar g_j^n(x)$ is less than 3\% in the relevant regions of $x$ and therefore Eq. (\ref{eq:gn}) can be useful in future analyses.}  
{In supplemental-material Sect. S3,     we  {checked that our results are consistent if the WW approximation for $g_2$ is used at the nucleon or at the nucleus level.}}
% quantity evaluated and discussed in Ref. \cite{CiofidegliAtti:1993zs}.  
 
 Encouraged by the successful test of the extraction procedure,  Eq. (\ref{eq:gn}) was applied to  the available data for $g_1^3(x)$ to get the corresponding neutron quantity.  The outcomes {(see section in supplemental material)} show that the model follows the extracted data for $g_1^n(x)$ and then plainly suggest that Eq. \eqref{eq:gn} provides a reliable method to get the neutron SSF. The differences between the Poincar\'e-covariant results and the ones based on the non relativistic polarizations are tiny with respect to the experimental errors, {that are expected to shrink in future EICs experiments.} 
 
{In closing, our Poincar\`e-covariant approach is  applied to establish  a robust baseline for more in-depth study of the Bjorken sum rule \cite{Bjorken:1969mm}. In fact, by calculating the first moment of the $^3$He SSF, $\Gamma_3$, one can extract the first moment of the neutron SSF, $\Gamma_n$, and proceed to evaluate the BSR, once $\Gamma_p$ is assumed well-known. As detailed   in supplemental-material Sect. S4, one finds a very nice agreement between the direct calculation of $^3$He first moment, obtaining $\Gamma_3=-0.0604$, and  the one evaluated by integrating the rhs of   Eq. \eqref{eq:gapp}  (with the nucleon polarizations listed in \eqref{eq:eff_pol}), that results to be $\Gamma_3^{pol}=-0.0611$.  Noteworthy, the tiny difference between the two values, $\sim 1$\% (cf. also the  quantitative comparison  in Fig. \ref{fig:3app}), motivates the application of a relation between first moments, analogous to Eq. \eqref{eq:gn}, in order to  extract $\bar \Gamma_n$. From the {GRSV parametrization}, at  $Q^2=10~GeV^2$, one gets  on one side $\Gamma^{GRSV}_n=-0.063$ and on the other $\bar \Gamma_n=-0.0622$, confirming the  validity of the extraction procedure. Summarizing, once  new measurements of $g_3(x)$ will be at disposal, one can apply the $\Gamma_n$ extraction, described above,   and provide a reliable estimate, based  on a careful description of  nuclear effects. }
 
% \begin{comment}
%   value of the Bjorken sum rule at $Q^2=10~GeV^2$ and the one where $\Gamma_n$, the first moment of $g_1^n(x)$, is evaluated through our $\Gamma_3$, is obtained within a theoretical uncertainty that results to be less than 3\%. This small value,  due to the approximation stemming from Eq. \eqref{eq:gn} (cf.  the  quantitative comparison  in Fig. \ref{fig:3app}),  
% \end{comment} 

 %Finally, in Fig. \ref{fig:dati} we compare our predictions for $g_1^3(x)$ (left panel) and $g_2^3(x)$ (right panel) with present data of Refs. \cite{E142:1996thl,JeffersonLabHallA:2003joy}. %Finally, one can see that the relativistic calculation, displayed as the full line on the right panel of Fig. \ref{fig:dati}, is in better agreement with the data w.r.t. the non-relativistic corresponding evaluation (dashed line in the same figure). 

{\it Conclusions.}
The covariant light-front approach, together with the Bakamajian-Thomas construction of the Poincaré generators, already successfully applied for investigating the EMC effect of  light nuclei \cite{Pace:2022qoj,Fornetti:2023gvf}, was  extended  to the calculation the $^3$He spin-dependent structure functions.
 To perform the evaluation, use was made of a very accurate nuclear wave function corresponding to the realistic  Av18 potential.
The dominance of the neutron contribution in $^3$He SSFs motivates the application of a workable procedure proposed in the past \cite{CiofidegliAtti:1993zs}, but with  the effective nucleon polarizations provided by the  Poincar\'e covariant framework. {On one side, the possibility of combining the refined description of the nuclear systems and the needed fulfillment of a fundamental symmetry, like the Poincar\'e covariance, yields a sound framework for the analysis of  the next generation of experimental  investigations {(we plan to use the present approach to evaluate polarized and unpolarized SF of heavier nuclei such as $^7$Li \cite{Li71,Li72} for which an experimental program is actually in preparation)}; on the other side, the present experimental accuracy does not allow to address  the relativistic effects, in a quantitative way. However,   the new extraction of $g^n_1(x)$ from available data, {shown in the supplemental material),}   agrees with both a previous extraction and the widely adopted GRSV parametrization  \cite{Gluck:2000dy}, without invoking any free parameter.  This confirms the robustness of the extraction procedure given in Eq. \eqref{eq:gn} 
 {allowing to extend  it to the first moment and gaining insights on the nucleon Bjorken sum rule}.} 
Therefore, the present investigation will be  {very useful} for the new generation of experiments involving polarized $^3$He nuclei,
such as those to be carried out at future EICs.

%{\it Acknowledgment.}
\begin{acknowledgements}
M.R. thanks for the financial
support received under the STRONG-2020 project of the
European Union’s Horizon 2020 research and innovation
programme: Fund no 824093.
\end{acknowledgements}
%%%%%% FIGURE

\newpage
\onecolumngrid

\centerline{\bf SUPPLEMENTAL MATERIAL}

\subsection*{S1: Nuclear hadronic tensor in IA}

%\replace{Here we discuss in details  how to properly express}
{This subsection is devoted to  details on the expression of} the antisymmetric nuclear hadronic tensor in terms of the one of the nucleon, %\replace{and hence how to parametrize it}
{as well as on its parametrimation}  in terms of the spin-dependent structure functions of the nucleons. Equation (7) of the main text can be rearranged as follows:
\begin{align}
     W^{a, \mu \nu}_A &=\sum_N ~\sumint {d\epsilon} \int \frac{d{\boldsymbol {\kappa}}_{\perp} ~d\kappa ^+} {2~(2 \pi)^3 ~  \kappa ^+}~ \frac{1} {\xi} 
     Tr \left [ {\boldsymbol{\hat{\mathcal{P}}}}^{N}(\tilde{\boldsymbol {\kappa}},\epsilon,{\textbf{S}},{\mathcal{M}})  ~{\boldsymbol{\hat{w}}}^{a,\mu \nu}_{N}({p},{q}) \right ] ~,
\end{align}
where ${\boldsymbol{\hat{\mathcal{P}}}}^{N}(\tilde{\boldsymbol {\kappa}},\epsilon,{\textbf{S}},{\mathcal{M}})$ and ${\boldsymbol{\hat{w}}}^{a,\mu \nu}_{N}({p},{q})$ are $2\times 2$ matrices.
In particular, the matrix ${\boldsymbol{\hat{w}}}^{a,\mu \nu}_{N}$ is:
\begin{equation}
    w^{a,\mu \nu}_{N,\sigma',\sigma}({p},{q})  = \frac{1} {2}  \left[{{\mathcal {W}}}_{N,0}^{{a,\mu \nu}}+ {\boldsymbol {\sigma}} \cdot {\boldsymbol {\mathcal{W}}}^{{a,\mu \nu}}_{N}({p},{q}) \right]_{\sigma'\sigma} ~,
\end{equation}
where: 
\begin{gather}
    {{\mathcal {W}}}_{N,0}^{{a,\mu \nu}} = Tr~ w^{a,\mu \nu}_{N,\sigma',\sigma}({p},{q}), \\
{\boldsymbol{\mathcal{W}}}^{{a,\mu \nu}}_{N}({p},{q}) = Tr ~\left [ {\boldsymbol {\hat {w}}}^{{a,\mu \nu}}_{N} ({p},{q}) ~ \boldsymbol {\sigma} \right ].
\end{gather}
Finally, the {vector} ${\boldsymbol{ \mathcal{W}}}^{{a,\mu \nu}}_{N}$  has components: 
\be 
     {\mathcal W}^{a\mu\nu}_{N,\ell}({p},{q})  = \frac{2 i ~ m \varepsilon^{\mu \nu}_{ ~~\rho \sigma} ~q^\rho} {p \cdot q } \Bigg \lbrace  S^\sigma_\ell  g^N_1 (z,Q^2) 
   +  \left [ S^\sigma_\ell - \frac{S_\ell \cdot q} {p \cdot q} p^\sigma \right ] g^N_2 (z,Q^2)   \Bigg \rbrace~~~~~~~~{\rm with}~(\ell=1,2,3)~,
\ee
%\replace{Here}
{where} the {four-}vector $S_\ell \equiv (S_\ell^0, S_\ell^1, S_\ell^2, S_\ell^3 )$ is:
\begin{align}
    S_\ell^0 &= \frac{p_\ell}{m}, \label{s0} \\
S_\ell^i &= \delta_{\ell i} + p_i ~ \frac{p_\ell} {m(p^0 + m)}, \quad  \quad \quad \text{with} ~ i=1, 2, 3 ~. 
\label{si}
\end{align}

\subsection*{S2: Explicit expression of the $^3$He SSFs}
%\replace{Here we show}
{This subsection illustrates} the complete expression of the { SSFs for a generic A-nucleon nucleus}, as functions of the nuclear TMDs,
$\Delta f^N (\xi, k_\perp^2), h_{1L}^{\perp N} (\xi, k_\perp^2), \Delta'_T f^N (\xi, k_\perp^2), h_{1T}^{\perp N}  (\xi, k_\perp^2)$, and $g^N_{1T}(\xi, k_\perp^2)$, already evaluated for $^3$He in Ref. \cite{Alessandro:2021cbg}. {One has}
\begin{align}\label{g1-fin}
    g_1^A(x)  = \frac{m^2 ~ 2 \pi}{M_A^2} \sum_N ~ \int  ~{ {k_\perp} d {k_\perp} ~ } \int^1_{\xi_{min}} d \xi ~ \frac{1}{\xi ^2} ~  \Bigg \lbrace g^N_1 (z)   &\left [ \Delta f^N (\xi, k_\perp^2) \left (  \frac{p_z}{m}  +  1 +  \frac{{p^2_z}}{m(p^0 + m)} \right  ) \right. 
\nonumber
\\
&+ \left.
  h_{1L}^{\perp N}  (\xi, k_\perp^2) ~  \frac{k_\perp^2}{M_A ~ m} \left  (  {1} + \frac{p_z ~}{(p^0 + m)}   \right ) \right ]  \Bigg \rbrace ~,
\end{align}
while:

\begin{align}
    &g_2^A(x)= \frac{m^2  ~ 2 \pi}{M_A^2}  ~  \sum_N ~ \int { {k_\perp} d {k_\perp} } \int ^1_{\xi_{min}} ~{d\xi }~ \frac{1 } {\xi^2 } ~ \left \{ g^N_1 (z) \right . \notag\\
 &\left .  \times   \left \{  \left [  \Delta'_T f^N (\xi, k_\perp^2) +  \frac{ k_\perp ^2   \left ( \Delta'_T f^N (\xi, k_\perp^2) + \frac{ k_\perp ^2} {2 M_A^2 } ~ h_{1T}^{\perp N} (\xi, k_\perp^2) \right )} {2 m (p^0 + m)}  + \frac{ k_\perp^2 ~ {p_z}  ~ g^N_{1T}(\xi, k_\perp^2)} {M_A m(p^0 + m)}  \right   ]  \right . \right .  \notag\\
&\left .  \left . -  \left [ \Delta f^N (\xi, k_\perp^2) \left ( \frac{p_z} {m}  +  1 +  \frac{p^2_z} {m(p^0 + m)} \right ) + h_{1L}^{\perp N}  (\xi, k_\perp^2) ~ \frac {k_\perp^2} {M_A m } \left  ( {1 } +  \frac{p_z}{(p^0 + m)} \right ) \right ] \right \} \right . \notag\\
&\left .  + ~  g^N_2 (z)  \right . \notag\\
 &\left . \times \left [  \Delta'_T f^N (\xi, k_\perp^2) + \frac{ {k^2_\perp} \left (  \Delta'_T f^N (\xi, k_\perp^2) + \frac{ k_\perp ^2} {2 M_A^2 } h_{1T}^{\perp N} (\xi, k_\perp^2) \right )} {2 ~ m} \left ( \frac{ 1} {(p^0 + m)} - \frac { 1 + \frac{~ p_z} {(p^0 + m)}} {p^+ }  \right ) \right .\right .  \notag\\
 &\left .  \left .  +  ~ \frac{1} {2 M_A}~ k_\perp^2 ~ g^N_{1T}(\xi, k_\perp^2) \left (\frac  {p_z}  {m(p^0 + m)} -\frac{  \frac{p_z} {m} + 1  + \frac {p^2_z}  {m(p^0 + m)}} {p^+ } \right )   \right ] \right \},
 \label{g2}
\end{align}
where:

\begin{align}
    p_z &=  \frac{1}{2} \left [  ( \xi -1) ~ M_A + \frac{ (A-1)^2m^2  + |{\textbf{k}}_{\perp}|^2} {M_A ~ (1 - \xi)}  \right ],
    \\
    p_0 &=  \frac{1}{2} \left [  ( \xi -1) ~ M_A + \frac{ (A-1)^2m^2  - |{\textbf{k}}_{\perp}|^2} {M_A ~ (1 - \xi)}  \right ]~.
\end{align}

In the above equations, terms of the order $\epsilon/m$ were neglected, with $\epsilon$ the energy of the 
interacting spectator  system. This approximation is allowed by the very rapid decrease of the nuclear spectral functions as the  values of $\epsilon$ increase. In the case of $^3$He, the spectral function becomes very tiny already for $\epsilon$ a few tens of MeV (see, e.g., \cite{PhysRevC.56.64}).

Let us remind that $\xi_{min}=x m/M_A$ and $z=mx/\xi M_A$.

Finally, as shown in Eq. (9) of the main text, the nuclear SSFs can be expressed as integrals over $\xi$ of the spin-dependent light-cone momentum distributions:
\be
\label{eq:gj}
 g_j^A(x)= g_j^{A,n}(x)+g_j^{A,p}(x)= \sum_{N=n,p} \int_{\xi_{min}}^1 \hspace{-0.2cm}d\xi ~
\Big\lbrace g_1^N(z) l^N_j(\xi)+ g_2^N(z) h^N_j(\xi)  \Big\rbrace~. 
%j=1,2
\ee

 By comparing the above equation with Eqs. \eqref{g1-fin} and \eqref{g2}, one gets:

\begin{align}
    h^N_1(\xi)&=0
    \\
    l_1^N(\xi)&=2 \pi \left(\dfrac{m}{M_A} \right)^2  \int dk_\perp ~k_\perp ~ \frac{1} {\xi ^2} 
    ~  \Bigg \lbrace   \Delta f^N (\xi, k_\perp^2) \left (  \frac{p_z} {m}  +  1 +  {{p^2_z} \over m(p^0 + m)} \right  ) +
  h_{1L}^{\perp N}  (\xi, k_\perp^2) ~  {k_\perp^2  \over M_A ~ m} \left  (  {1} + {p_z ~  \over (p^0 + m)}   \right )  \Bigg \rbrace ~,
  \label{eq:l1}
  \\
  h_2^N(\xi) &= 2\pi \left(\dfrac{m}{M_A} \right)^2  \int dk_\perp ~k_\perp {1 \over \xi ^2} ~\left [  \Delta'_T f^N (\xi, k_\perp^2) + { {k^2_\perp} \left (  \Delta'_T f^N (\xi, k_\perp^2) + {  k_\perp ^2 \over 2 M_A^2 } h_{1T}^{\perp N}  (\xi, k_\perp^2) \right )  \over 2 ~ m} \left  (   { 1 \over (p^0 + m)}  -{  {1 }  +   \frac{~ p_z} {(p^0 + m)} \over p^+ }  \right ) \right . \notag\\
 &\left .    +  ~ \frac{1} {2 M_A}~ k_\perp^2 ~ g^N_{1T}(\xi, k_\perp^2) \left ( ~ \frac{p_z} {m(p^0 + m)} - \frac{ \frac{p_z} {m} + 1  +  \frac{p^2_z} {m(p^0 + m)}} {p^+ } \right )   \right ] ,
 \\
 l_2^N(\xi) &= 2\pi \left(\dfrac{m}{M_A} \right)^2
 \int dk_\perp ~ k_\perp \frac{1} {\xi ^2} ~  \left \{ \left [ \Delta'_T f^N (\xi, k_\perp^2) +  \frac{ k_\perp ^2   \left ( \Delta'_T f^N (\xi, k_\perp^2) + \frac{ k_\perp ^2} {2 M_A^2 } h_{1T}^{\perp N}  (\xi, k_\perp^2) \right )} {2 m (p^0 + m)}  + \frac{ k_\perp^2  {~ {p_z} } ~ g^N_{1T}(\xi, k_\perp^2)} {M_A m(p^0 + m)}  \right   ]  \right .  \notag\\
&  \left . -  \left [ \Delta f^N (\xi, k_\perp^2) \left ( \frac{p_z} {m}  +  1 +  \frac{p^2_z} {m(p^0 + m)} \right ) + h_{1L}^{\perp N}  (\xi, k_\perp^2) ~  \frac{k_\perp^2} {M_A m } \left ( {1 } + \frac{p_z} {(p^0 + m)} \right ) \right ] \right \}~.  \notag
  \label{eq:l2}
\\
\end{align}
It is worth recalling that the support of the spin-dependent light-cone distribution is automatically fulfilled within the light-front Hamiltonian dynamics framework we are exploiting, i.e. without any ad hoc  constraint.
{Let us introduce further details on the approximation Eq. (10) of the main text:}
\begin{align}
\bar g_j^A(x)  = 2 p^p_j g_j^p(x)+p^n_j g_j^n(x)~ 
\label{eq:l3}
\end{align}
{As a matter of fact the nuclear dynamics leads to a nucleon momentum distributions peaked at $|\vec k|=0$ (being $\vec k$ the nucleon momentum in Cartesian variable). As a consequence, the TMDs are strongly peaked around $\xi \sim 1/3$ and the contribution of $p_z$ and $p_z^2$ are  negligible in  Eqs. (\ref{eq:l1})-(\ref{eq:l2}). Moreover, terms like 
 $k_\perp^2/m^2$  are negligible with respect to 1 in  Eqs. (\ref{eq:l1})-(\ref{eq:l2}). Therefore, the corresponding light-cone distributions are peaked around $\xi \sim 1/3$ too. Then one obtains Eq. (\ref{eq:l3}). In Fig. \ref{fig:app} we compare the full calculation of, e.g., $g_1^3(x)$ with $\bar g_1^3(x)$. As one can see, differences are extremely small {up to $x=0.9$}}.

\begin{figure*}[h]
\begin{center}
\includegraphics[width=8.8cm]{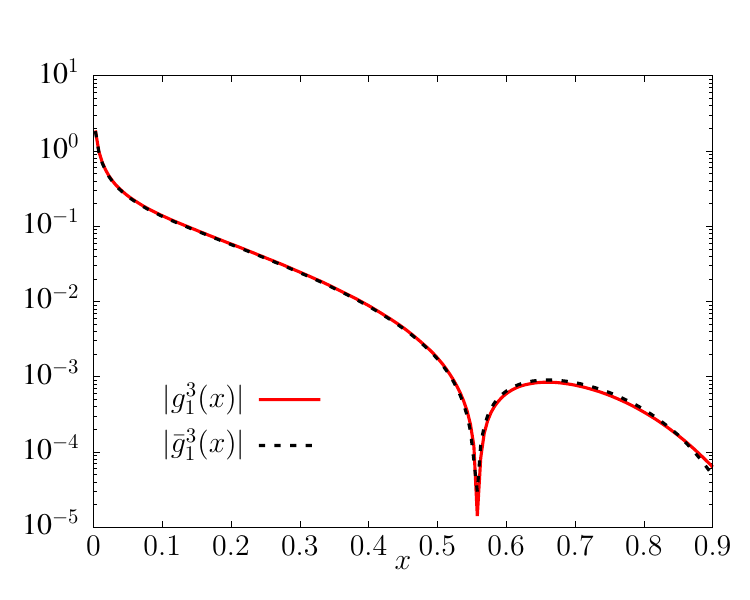}
\caption{ (Color online). 
{Comparison between $|g_1^3(x)|$ (full line) and $|\bar g_1^3(x)|$ (dashed line). }
 }
\label{fig:app}
\end{center}
\end{figure*}

{Before closing, in order to deeply test the validity of the present approach, we compare in Fig. \ref{fig:x2g1g2} the calculation of $x^2 g_1^3(x)$ and $x^2 g_2^3(x)$ {using Eq. (\ref{eq:gj})} with data \cite{E142:1996thl,JeffersonLabHallA:2003joy,JeffersonLabHallA:2016neg}.  Also in this case the calculations are in fair agreement with the data.}
{In closing this section, let us show how the above Eq. (\ref{eq:l3}) can be used to experimentally extract the neutron SSFs from $^3$He data:}

\be
\label{eq:gn1}
    \bar g_1^n(x)= \dfrac{1}{p_1^n} \big[g_1^3(x)- 2 p_1^p g_1^p(x)  \big]~,
\ee
In Fig. \ref{fig:n_extr}, the results of the extraction procedure applied to the experimental data, shown in the left panel of Fig. {1 of the main text}
%\ref{fig:g1g2} 
(taking into account the error bars), are compared to i)  the GRSV parametrization  \cite{Gluck:2000dy} and ii) previous extraction discussed in Ref. \cite{E142:1996thl}.

\begin{figure*}[h]
\begin{center}
\includegraphics[width=8cm]{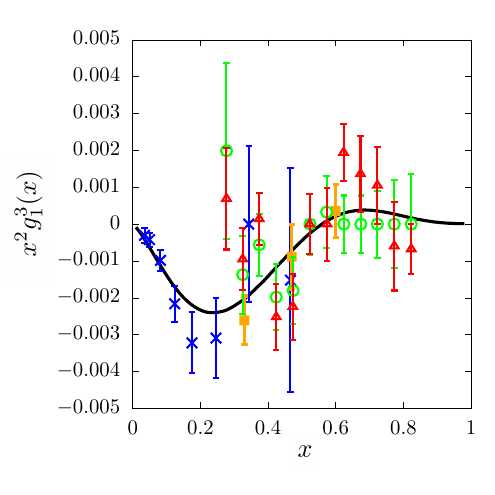}
\includegraphics[width=8cm]{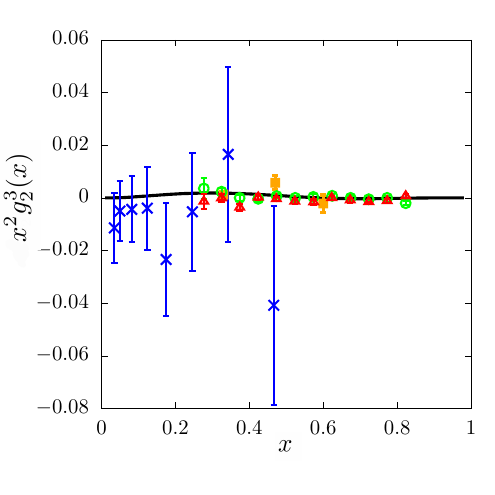}
\caption{ Color online). The $^3$He SSFs obtained with the Poincaré covariant approach (full line)  vs. the Bjorken variable $x$. The experimental data are taken from 
Ref. \cite{E142:1996thl} (crosses), Ref. \cite{JeffersonLabHallA:2003joy} (squares), and Ref. \cite{JeffersonLabHallA:2016neg}:  {empty circles} correspond to an incoming-electron energy ${\cal E}=5.89$ GeV and  triangles   to ${\cal E}= 4.74$ GeV). 
Left panel:  $x^2 g^3_1(x)$. Right panel:   $x^2 g^3_2(x)$.  }
\label{fig:x2g1g2}
\end{center}
\end{figure*}

\begin{figure}[t]
\begin{center}
\includegraphics[width=8.0cm]{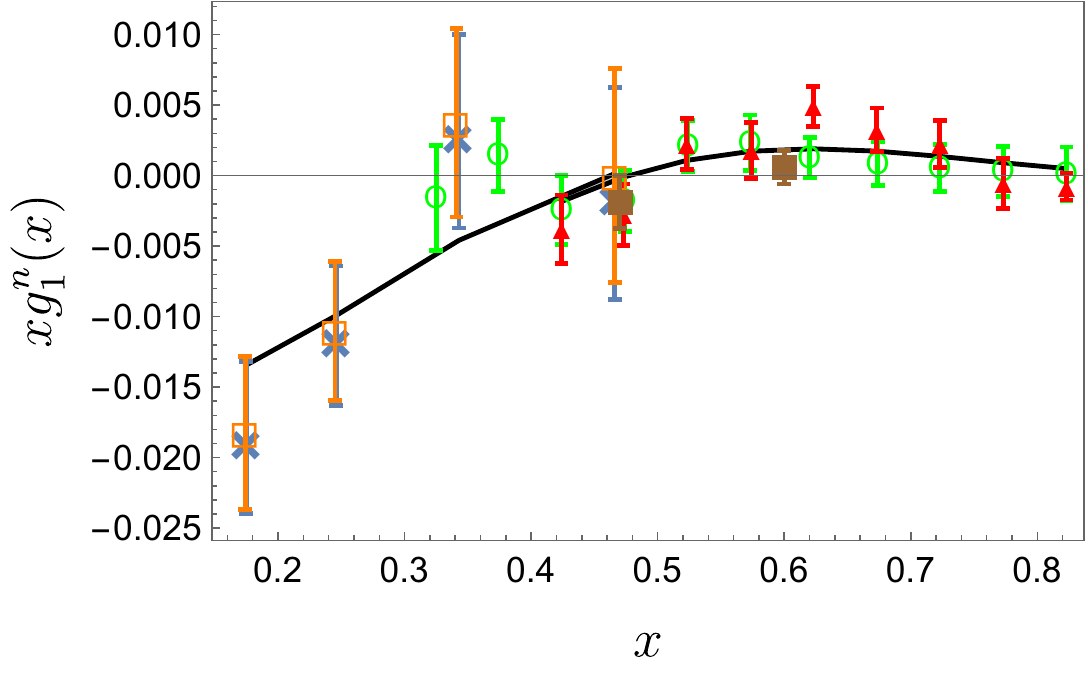}
\caption{ (Color online). 
{$x \bar g_1^n(x)$ extracted by using Eq. (\ref{eq:gn1})} and  the $^3$He data for   corresponding quantity. Full line is  the neutron SSF from the GRSV parametrization   \cite{Gluck:2000dy}. {Symbols correspond to those in Fig. 5 of the SM}.
 The empty squares are the  neutron SSF obtained in Ref. \cite{E142:1996thl}.  } 
\label{fig:n_extr}
\end{center}
\end{figure}

\subsection*{S3: Test of the Wandzura-Wilczek
relation for the $^3$He}
{
In this subsection, we present the test of coherence of the WW approximation,  for both  $^3$He  and  nucleon, within our framework.  To this end: i) in the $^3$He case,
we compare $g_2^3(x)$ evaluated by Eq. \eqref{g2} with the  value obtained from the WW approximation  through $g_1^3(x)$ obtained from   Eq. \eqref{g1-fin}; ii) in the nucleon case, we compare the neutron SSF $\bar g_2^n(x)$ obtained through Eq. (13) of the main text with $j=2$, i.e. by using $g_2^3(x)$ evaluated by Eq. \eqref{g2} and the effective polarizations, with the result
%$g_2^n(x)$ 
obtained by applying the WW approximation to $\bar g_1^n(x)$, also extracted from the same  Eq. (13) but with $j=1$.
In Fig. \ref{fig:WW} we display the results of the above comparisons.}
{In both cases, the WW approximation predicts similar distributions if applied at both  nuclear and  nucleonic levels. In particular, the main differences can be found near the second minimum where the SSF changes the sign and at high $x$, i.e. in the regions where the SSFs are small. }

\begin{figure*}[h]
\begin{center}
\includegraphics[width=8.8cm]{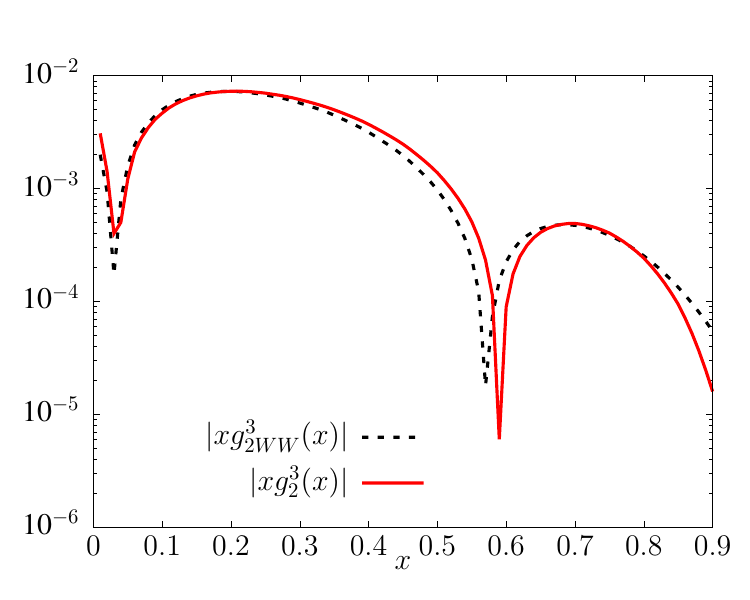}
\includegraphics[width=8.8cm]{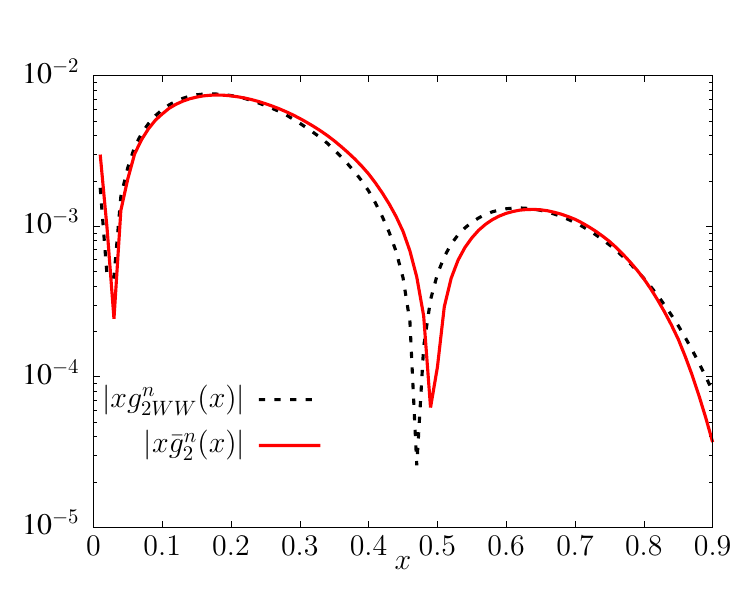}
\caption{ (Color online). 
{Left panel:  the full line represents the calculation of the $x g_2^3$ SSF where use was  made of Eq. (\ref{g2}). Dashed line, same quantity obtained by applying the WW approximation to $x g_1^3$, i.e. at the nuclear level. Right panel: the full line is the extracted neutron SSF $x \bar g_2^n$ (result shown in the right panel of Fig. 4 of the main text). Dashed line, the same quantity but the SSF was obtained by applying the WW approximation to  $x \bar g_1^n$.  }
 }
\label{fig:WW}
\end{center}
\end{figure*}

\subsection*{S4: Test of the Bjorken sum rule}

In this subsection,  the Bjorken sum rule (BSR) \cite{Bjorken:1969mm}, which is directly connected to the first moment of   SSFs, is discussed. To this end, let us first define the non-singlet quantity $\Gamma^{NS}_1$:
\begin{align}
    \Gamma^{NS}_1 \equiv \Gamma_p-\Gamma_n~,
\end{align}
where in general {the definition} $\Gamma_A = \int_0^1 dx~ g_1^A(x;Q^2)$ {can be introduced} for both nucleons and nuclear systems.
 At finite $Q^2$, taking into account the pQCD corrections, the BSR reads
\begin{align}
  \Gamma^{NS}_1 = {\rm BSR} = \dfrac{1}{6}\dfrac{g_A}{g_V} C(Q^2)~,
    \label{eq:bjM2}
\end{align}
with $(1/6)~g_A/g_V \sim 0.2126 \pm 0.0002$ \cite{Workman:2022ynf}. The function $C(Q^2)$, that accounts  for the finite-$Q^2$ corrections, is given at NNLO and in the minimal-subtraction scheme $\overline{\rm MS}$, by  the following expression (see, e.g., Refs. \cite{Larin:1991tj,Deur:2018roz})
\be
C(Q^2)= 1- {\alpha_S^{\overline{\rm MS}}(Q^2)\over \pi}+\Bigl({n_F\over 3}-{55\over 12}\Bigr){[\alpha^{\overline{\rm MS}}_S(Q^2)]^2\over \pi^2}~.
\ee 
For instance, by using the  {NLO parametrizations  of Tab. I in Ref. \cite{Gluck:2000dy}  (GRSV) at the  scale $Q^2\sim 10~GeV^2$, one gets   $\Gamma^{GRSV}_p - \Gamma^{GRSV}_n=0.196$, with  $\Gamma^{GRSV}_p=0.133$ and $\Gamma^{GRSV}_n=-0.063$  (cf.   Tab. II of Ref. \cite{Gluck:2000dy}).} This result should be compared to   Eq. \eqref{eq:bjM2}, that at NLO gives $BSR=0.199$, once  $\alpha_S^{\overline{\rm MS}}(Q^2=10~GeV^2) \sim 0.25$  is taken. The difference could point to the need of a better determination of $\Gamma_n$, once new parametrizations of the neutron SSFs will be available.

\begin{comment}

To be specific, by still keeping $\Gamma_p=0.133$, as given in Ref. \cite{Gluck:2000dy}, one should have $\Gamma^{BSR}_n=-0.066$, which is different from $\Gamma_n=-0.062$, given in  Ref. \cite{Gluck:2000dy}. In what follows, we illustrate the possibility to extract $\Gamma_n$ from the correspond $^3$He quantity, $\Gamma_3$, assuming  and testing that $\Gamma_3=p_1^n\Gamma_n + 2 p_1^p \Gamma_p$, as one could argue from  Eq. (11) in the main text. First of all, the comparison between a direct calculation of $Gamma_3$, within our Poincar\'e framework, yields $$\Gamma_3 = -0.0535 \pm  0.0005 ~,$$
for $Q^2=10~GeV^2$, while $$p_1^n\Gamma_n + 2 p_1^p \Gamma_p \sim  -0.059~,$$ by using  the nucleon polarized SSFs of Ref. \cite{Gluck:2000dy}, in both sides. Then 
\be 
\Gamma_n={\Gamma_3-2p^p_1 \Gamma_p\over p^n_1}= -0.0543\pm  0.0006
\ee
with $p^n_1\simeq 0.873$ and  $p_1^p \simeq -0.0230$, as given in the main text. Within the above approximation, the calculated BSR is $\Gamma_1^{NS}(th)=0.1873$. While the $^3$He SSF $g^3_1(x)$ is well approximated by the expression $p_1^n g^n_1(x) + 2 p_1^p g^p_1$, with an uncertainties of the order 3-5\%, while  the approximation works at 10\% for $\Gamma_3=p_1^n\Gamma_n + 2 p_1^p \Gamma_p$. \red{Questa differenza andrebbe capita...}. A better parametrization of $g^n_1(x) $, in particular at small $x$, driven by new accurate data, could contribute to reduce the discrepancy  between expected BSR and calculatred one.
\end{comment}

{Indeed,  one could  exploit in full $g^3_1(x)$.} The procedure presented in the main text, which aims to extract the neutron SSFs from the corresponding proton and $^3$He data (see the main-text Eq. (13)), can be extended to get $\Gamma_n$ from $\Gamma_3$, given by
\begin{align}
    \Gamma_3 \equiv \int_0^1 dx~ g_1^3(x)~.
\end{align}
From the main-text Eq. (13), $\Gamma_3$ can be approximated by \begin{align}
    \Gamma_3^{pol} \equiv p_1^n \Gamma_n + 2 p_1^p \Gamma_p~,
\end{align} 
where $p_1^p$ and $p_1^n$ are the effective polarizations defined in the main-text Eq. (12). { At $Q^2=10~GeV^2$,  using  the GRSV NLO results for $\Gamma^{GRSV}_{p(n)}$,  one gets }  $\Gamma_3^{pol}=-0.0611$.  Differently, from our Poincar\'e-covariant evaluation of $g_1^3(x)$, one has $\Gamma_3 = -0.0604$.  {Such a  difference between the two outcomes,    basically } due to nuclear effects not fully accounted  for by $p_p$ and $p_n$,  is  quite tiny,  $\sim 1.2\%$.  
Therefore, thanks to this successful comparison, one can  extract $\Gamma_n$ from $\Gamma_3$, once   $\Gamma_p$ is known, as follows
\begin{align}
    \bar \Gamma_n \equiv \dfrac{1}{p_1^n} \Big[\Gamma_3-2 p_1^p \Gamma_p  \Bigr]~.
    \label{eq:gnex}
\end{align}
From our above result for $\Gamma_3$ (recall that in $g^3_1(x)$ the NLO nucleon SSFs of Ref. \cite{Gluck:2000dy} are adopted), one gets $\bar \Gamma_n \sim -0.0622$, that agrees with  $\Gamma^{GRSV}_n$  within a $1.3 \%$ error. {This crosscheck  validates the proposed extraction of $\Gamma_n$, once an accurate experimental $g^3_1(x)$ is provided and a reliable estimate of $\Gamma_3$ becomes available.
{Before closing this section, let us point out {that} the theoretical error in estimating $\Gamma_3$ is due to {{ the numerical treatment of the integrable divergence of $|g_1^p(x)|$ for $x \rightarrow 0$}.  
} }

\bibliography{emc.bib}

\end{document}